# DON'T LOOK BACK: AN ONLINE BEAT TRACKING METHOD USING RNN AND ENHANCED PARTICLE FILTERING

*Mojtaba Heydari and Zhiyao Duan[1]*

Department of Electrical & Computer Engineering
University of Rochester, 500 Wilson Blvd, Rochester, NY 14627, USA
mheydari@ur.rochester.edu, zhiyao.duan@rochester.edu

**ABSTRACT**

Online beat tracking (OBT) has always been a challenging task. Due to the inaccessibility of future data and the need to make inference in real-time. We propose *Don't Look back!* (DLB), a novel approach optimized for efficiency when performing OBT. DLB feeds the activations of a unidirectional RNN into an enhanced Monte-Carlo localization model to infer beat positions. Most preexisting OBT methods either apply some offline approaches to a moving window containing past data to make predictions about future beat positions or must be primed with past data at startup to initialize. Meanwhile, our proposed method only uses activation of the current time frame to infer beat positions. As such, without waiting at the beginning to receive a chunk, it provides an immediate beat tracking response, which is critical for many OBT applications. DLB significantly improves beat tracking accuracy over state-of-the-art OBT methods, yielding a similar performance to offline methods.

## 1. INTRODUCTION

Beat tracking is an important task in Music Information Retrieval (MIR) and is the core block of music rhythm analysis. It also has a vast range of usages for other MIR tasks (e.g., music transcription, music metadata generation, genre classification, music emotion recognition, and event analysis). Online beat tracking (OBT) refers to beat tracking in real time or a causal fashion. It enables further applications such as automatic music accompaniment, interactive music processing, and auto DJs.

In general, every music beat tracker has two main blocks. One is extracting some music features that are informative for the task and the second one is to infer beat positions based on the extracted features. For the first block, many approaches [1-2] extract an onset strength signal (OSS) such as signal's spectral flux as the main feature for beat detection. Other methods [3-4-5] take advantage of deep neural network to directly obtain a beat activation function. With the OSS or beat activation function, in the second block, different techniques such as autocorrelation function (ACF) and peak picking [4], resonating comb filter (CF) [24], dynamic programming [8] or more advanced probabilistic models such as hidden Markov models and dynamic Bayesian networks (DBN) [25], Kalman filter [15], and particle filtering (PF) [17-18] are used to infer beat locations.

Compared to offline approaches, OBT faces unique challenges including partial data access, processing speed constraints, and the inability to correct previous detections. One category of existing OBT methods use a moving window to model signal regularity over the recent past signal to make beat predictions [7-8-9-10-11-13-14]. In particular, OBTAIN [6] uses cumulative OSS and autocorrelation response to locate beats. However, the cumulative OSS relies heavily on past data and cannot track tempo changes or fluctuations properly. Aubio [13-14] locates beat positions based on the weighted autocorrelation response in a context-dependent model of OSS. Böck and Schedl [7] used beat activations from a recurrent neural network (RNN) with a bidirectional long short-term memory (BLSTM) structure in a moving window to estimate local tempo and infer beats based on auto correlation function. Gkiokas and Katsouros [8] utilized the activation functions of a CNN and dynamic programing for a moving window to estimate next beat positions. There are several arguments against the moving window strategy. These include the potential for computational overload, the intrinsic difficulty in adapting these tracking strategies to causal and real-time scenarios, and the lack of continuity between windows [12].

There are a few OBT methods that do not use a moving window. IBT [12] is an OBT model based on the Beatroot [2] multi-agent offline approach, which infers beat positions without using a dynamic window. However, the initialization of the agents can be difficult, needing up to 5 seconds of audio. Shiu et al. [15] utilized Kalman filtering for beat tracking, but the underlying Gaussian assumption of the observation distribution can be too strong.

In contrast, particle filtering (PF) is a more general nonparametric approach for sequential decision-making problems. Duan and Pardo [16] used PF for online audio score alignment. Cemgil and Kappen [17] applied PF in tempo tracking and rhythm quantization given a prior on quantization locations. Hainsworth and Macleod [18] used it in tempo tracking. According to [17-18], a critical component of PF methods is the choice of appropriate input features and the related observation models. Another practical consideration is reducing the number of particles to speed up inference for real-time applications.

---
[1] Thanks to National Science Foundation Grant 1846184.

In this paper, we propose a novel OBT model that integrates a unidirectional RNN for feature extraction and particle filtering for online decision making. In particular, the RNN predicts a beat activation function for each incoming audio frame following [25]. For the particle filtering part, we utilize an efficient state space and transition model [21], which speed up the process and reduce the need to use a great number of particles. We also propose a new observation model. We further propose informative priors to improve the performance. Experiments on the GTZAN dataset show that our proposed model outperforms state-of-the-art methods and can be utilized in situations with weaker processing capacity.

## 2. APPROACH

In this section, we describe our DLB approach to music OBT. The name of the model reflects the fact that it does not require the beat activations of past frames to make an inference about the current frame. Here are two points to clarify though. 1) No need to look back is an intrinsic property of PF models, not limited to our system. However, we highlight this fact because such a characteristic can be advantageous for OBT applications. 2) By saying no need for past activations, we specifically refer to the inference stage that unlike moving window approaches, it does not need previous activations. We do recognize that activation generation process itself uses an RNN to implicitly model long-term dependencies in past data.

### 2.1. Pre-processing

RNN structures have been an interesting choice for many time series applications, since they consider the relationship between adjacent frames of data. In particular, many recent works in related fields take advantage of RNNs with BLSTM neurons. The main advantage of BLSTM over LSTM is its better performance, which originates from leveraging future data consideration in addition to past data. However, given that causality (which makes the future data inaccessible) is one of the main conditions in OBT applications, we chose to use an LSTM neural net trained to extract beat activations as the input of our PF inference model. In the preprocessing module, a Hann window is applied to the current frame of the audio signal with a size of 46 ms and hop size of 10 ms. Then, after obtaining the magnitude spectrogram, a logarithmically spaced filters ranging from [30, 17k] Hz with 12 bands per octave, corresponding to semitone resolution is applied. The first-order temporal differences are concatenated to the log-scaled spectrogram and then, the total spectrogram is fed to an ensemble of 8 pretrained LSTM nets. Their output is averaged to get current frame's beat activation. Note that to obtain the activation from the ensemble of LSTMs, pretrained models obtained from Madmom python library [19] are used. After getting the activation of the current frame, we feed it to our PF inference block to classify it to either the beat or the non-beat class.

### 2.2. Inference

In this paper, we approach the problem from the PF localization perspective, which comprises of two steps of motion and correction. According to the importance sampling principle, a high dimensional probability distribution $p(x) = p'(x)/C$ with an unknown model, parameters, and normalization constant $C$, can be represented by a large number (N) of independent samples $x^{(i)}$ from a known arbitrary proposal distribution $\pi(x)$. Even though there is no strict limitation in choosing $\pi(x)$, choosing a closer one to the distribution of interest, leads to a more efficient representation of $p(x)$. Thus:

$$p(x) = \frac{p'(x)}{C} = \frac{p'(x)}{C\,\pi(x)}\pi(x) = \frac{p'(x)}{C\,\pi(x)} \lim_{N\to\infty} \frac{1}{N}\sum_{i=1}^{N} \delta(x - x^{(i)})$$

$$= \lim_{N\to\infty} \sum_{i=1}^{N} \frac{\omega_i}{\sum_{i=1}^{N}\omega_i}\, \delta(x - x^{(i)}) \quad (1)$$

Where $\omega_i = p'(x^{(i)})/\pi(x^{(i)})$ is each particle $i$'s importance weight.

Now, consider the unknown distribution of interest $p(x_k|y_k)$ in our decoding problem, where $x_k$ and $y_k$ are beat hidden states and the signal's observations at each step $k$, respectively. Now, by considering Bayes filter, chain rule for the 1st order Markovian process, and assigning the proposal distribution equal to states transition probability, the problem is boiled down to a sequential Monte Carlo posterior calculation as follows:

$$\pi(x_{0:K}^{(i)}|y_{0:K}) = \prod_{k=1}^{K} p(x_k^{(i)}|x_{k-1}^{(i)})\, p(x_0^{(i)}) \quad (2)$$

$$p(x_{0:K}^{(i)}|y_{0:K}) \propto \prod_{k=1}^{K} p(y_k|x_k^{(i)}) p(x_k^{(i)}|x_{k-1}^{(i)}) \quad (3)$$

$$\omega_K^{(i)} = \frac{p'(x_{0:K}^{(i)}|y_{0:K})}{\pi(x_{0:K}^{(i)}|y_{0:K})} = \frac{p(y_K|x_K^{(i)})\, p(x_K^{(i)}|x_{K-1}^{(i)})}{p(x_K^{(i)}|x_{K-1}^{(i)})}\, \omega_{K-1}^{(i)} \Rightarrow$$

$$\omega_K^{(i)} = p(y_K|x_K^{(i)})\, \omega_{K-1}^{(i)} \quad (4)$$

Formulae (4) states that if we sample from a transition distribution (motion) as a proposal distribution of $\pi$, we can update the weights simply by multiplying them with the likelihood of the current frame. Thus, to infer beat positions using the posterior (1) estimation for each frame, the following algorithm will be used:

*1- Sample particles from proposal distribution (2) which in our case is transition probability. (**motion**)*
*2- Compute the new importance weights (4) based on observation probability derived from LSTM Beat activations.*
*3- Resample based on new normalized weights that discard unlikely hypotheses and generate more rational ones. (**correction**)*
*4- Take the median of the positions of all particles and classify the frame as beat if it is within the beat boundary and is far from previous beat with a dynamic time threshold equal to half of the median of all particle's tempo.*

### 2.3. State space and transition model

One popular state space model for such problems is the bar pointer model [20] which jointly models tempo and position within the bar as hidden variables. However, for the state space and transition model in this paper, rather than using the common bar pointer model, we implemented the model that is described in [21]. According to that, at each time frame k, the hidden state is referred to as $x_k=[\Phi_k, \dot\Phi_k]$, with $\Phi_k \in \{1,2,\ldots,M\}$ denoting the position within the bar and $\dot\Phi_k \in \{\dot\Phi_{min}, \dot\Phi_{min}+1,\ldots,\dot\Phi_{max}\}$ denoting the jumping intervals representing tempo of the current state. State space and transition probability are then defined as follows:

$$M = \lfloor \frac{60}{T\times\Delta} \rfloor \quad (5)$$

$$p(\dot{\Phi}_k|\dot{\Phi}_{k-1}) = \begin{cases} exp(-\lambda \times |\frac{\dot{\Phi}_k}{\dot{\Phi}_{k-1}} - 1|) & for\ \Phi_k \in B \\ \dot{\Phi}_k = \dot{\Phi}_{k-1} & for\ \Phi_k \notin B \end{cases}, \quad (6)$$

Equation (5) relates each tempo in BPM denoted by T to its corresponding integer jumping interval M, and Δ denotes frame hop size in seconds. Each row of the 2D state space shown in Fig.1 represents a different M. Note that we model beat state space rather than bar state space. Equation (6) is the transition probability where $\lambda$ is the parameter to decide the sampling width, B is beat states collection (yellow dots in Fig. 1). To specify the B boundaries that determine beats/non-beat states, we propose three models as shown in Fig. 1. Model (a) sets a vertical line as beat/non-beat boundary. Model (b) considers a constant number of states at each row as beat states. Model (c) provides a smooth Gaussian transition between $b_k$ and $\gamma$.

We believe that our described model is very suitable for PF applications for two main reasons: 1- It needs fewer number of states compared to the original beat pointer model (→fewer number of particles can represent the state). 2- Allowing tempo shifts only in beat positions lowers computational cost. Instead of sampling in every step, we do it only in beat positions (B states), and for the rest of states, given that the tempo is the same, we only shift particles. Considering the transition model, to detect tempo changes, no jitter or particle noise is required. For double/half tempo investigation, some works add a Metropolis-Hastings step to the algorithm. However, it was found in [18] that this step was counterproductive. In this paper we tried two different models for double/half tempo investigation. One was adding some probability terms to $\dot{\Phi}_k/2$ and $\dot{\Phi}_k \times 2$ in the first line of the transition probability (6). However, since the transition probability affects all the particles, it caused the model to diverge. The other method was adding double/half tempo investigation only to the median of tempi of all particles (interpreted as current frame's tempo). So, we feed the information of the median of tempi to the resampling block as informative priors.

### 2.3. Observation model

In this paper, we present a new observation model in which, localization is conducted only using the presence of landmarks as follows:

$$p(y_k|x_k) = \begin{cases} b_k & x_k \in B \\ \gamma & x_k \notin B \end{cases}, \quad (7)$$

where $b_k$ is the beat activation of frame k, and $\gamma$ is a small constant number (e.g., $\gamma = 0.03$). This means that in our approach, rather than using non-beat activations (which are the second output of the softmax in the RNN and fixed to be complementary to the beat activations) as clues in the likelihood function, we only rely on beat activations as landmarks and set all non-beat states probabilities to a small constant number. This is because the softmax output of the RNN is the posterior probability of beat/non-beat instead of the likelihood. Our experiments also showed that low beat activation (→high non-beat activation) does not necessarily mean that the frame is not a beat frame. In many cases, due to reasons like low percussive content or low frame energy, beat activations are weak. In traditional approaches, the complement of these activations encourages the model to classify some of the beat frames as non-beat [21-25]. Figure 2 demonstrates the whole PF inference process. A demo of the performance of our model can be found in following footnote[2].

---
[2] https://www.youtube.com/watch?v=u2Ee6WsNzoU

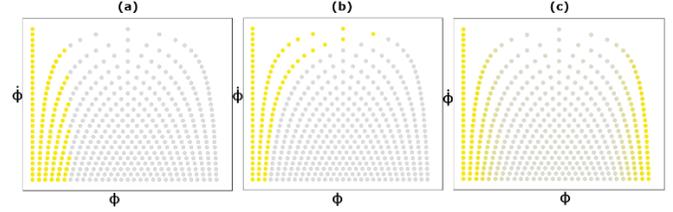

**Fig. 1:** Different beat/non-beat boundary models in the state space. (a): fractional beat/non-beat discriminator (b): constant number discriminator. (c): Gaussian soft discriminator.

## 3. EVALUATION

In order to evaluate the performance of our online model, we compare it with state-of-the-art online methods as well as state-of-the-art offline methods.

### 3.2. Evaluation metrics

In line with most other beat tracking publications, F-measure ($F_1$) with a tolerance window of ±70 (ms) is reported as the main performance criteria. To investigate the required time for each method to become initialized, we evaluate with and without discarding the predictions of the first 5 seconds of each recording.

### 3.2. Dataset

We used GTZAN dataset [22-23], a large dataset containing 1000 music pieces with a duration of 30 seconds each covering 10 different music genres (e.g., blues, classic). We used the entire dataset as the test set to illustrate the performance of different methods in a more general fashion. It is important to note that this dataset is not used to train the RNN.

### 3.3. Evaluation results

The DLB model does not require informative priors in the initialization step. Therefore, to initialize early particles we used a random uniform distribution that gives the same chance of receiving a hypothesis to each state. N=1000 is the number of particles that are used in Table 1 evaluations. The transition parameter is λ=30, and for the observation model, the fractional discriminator with the parameter of 1/60 is used. As shown in table 1, for both 5SecSkip and NoSkip, The DLB method outperforms all state-of-the-art online beat tracking models. Comparison between two $F_1$ columns in table 1 provides insights about the performance quality of different methods at the beginning, comparing to their general performance. Shifting from 5SecSkip to the NoSkip setting, the performance of ACF and IBT models decrease dramatically, while DLB and Aubio decline mildly. However, one important point to mention is that the IBT model's small number of agents must be initialized using the first 5 seconds of audio. It does not deliver any output during this time period. Similarly, the ACF model needs to wait for 6 seconds at the beginning to calculate the autocorrelation function to estimate beat positions. Note that in this evaluation, ACF is implemented in a causal fashion. Also, the lowest performing deference with and without skip belongs to Aubio. The reason is that unlike the DLB model, which initializes particles randomly and delivers results immediately (figure.2 a), Aubio is more patient and by considering relations between a few early beats, comes up with more careful results at the beginning.

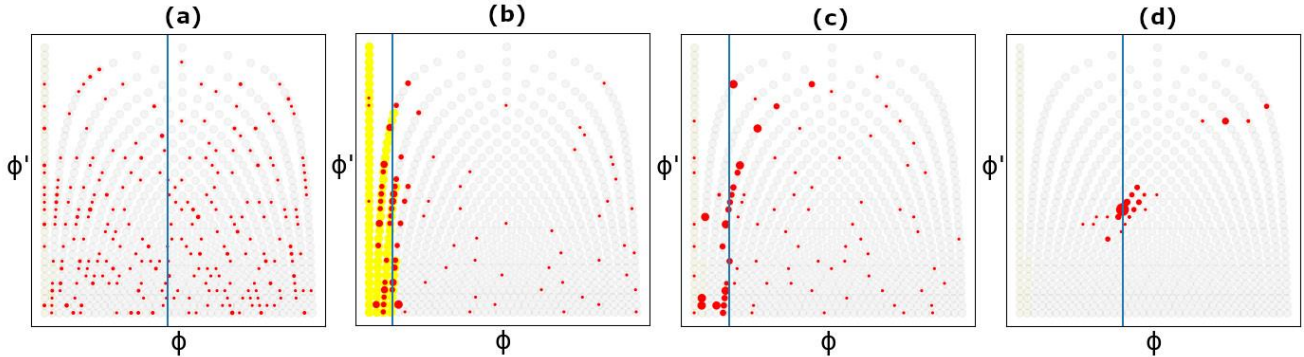

**Fig. 2:** Proposed PF inference process. (a): particles are initialized randomly and start to move right one step per frame (b): particles within the beat boundary gain weight while many others get discarded, when the first strong beat activation arrives. (c): significant gatherings move right with different paces. (d): Upon the next beat activation's arrival, many gatherings are discarded and the one with the correct tempo survives; a few double tempo investigators are also added. Blue line is the median of particles' positions.

In addition to online models, in the remainder of Table 1, performances of two offline methods are reported as well. The objective of considering offline methods is to compare between their inference and that of particle filtering. The compared offline models are resonating comb filters [24] and a dynamic Bayesian network [5], which is the state-of-the-art for offline beat tracking. It is important to note that even though the original offline models used BLSTM neurons in their RNN structure (which led to better performance in offline fashion), we replaced them with LSTM neurons in order to draw a fair comparison between DLB inference and those inferences. So, the input of all three models is common. Even considering DLB's causal setup, its performance is close to the non-causal DBN model. In addition to performance, the other very important factor in real-time usage is processing speed. PF models are slow in general. However, since our model uses an efficient state and transition model which only requires resampling in beat positions (for the rest of states, particles just shift forward by one step per frame) and uses stochastic universal sampling in the correction block (which only needs one random number generation rather than N random numbers for N particles), then our model using the default setting of N=1000 is already works in real time. However, to make it usable in situations with weaker processing capacity (e.g. microcontrollers), it is better to use a smaller number of particles.

**Table 1:** F-measure report of online/offline beat tracking models and initialization time for online models (GTZAN dataset).

| | Online beat tracking methods | | |
|---|---|---|---|
| Method | F-measure (5 Sec. skip) | F-measure (without skip) | Initialization elapse Time |
| **DLB** | 73.77 | 71.44 | 0 seconds |
| ACF [7] | 64.63 | 51.74 | 6 seconds |
| IBT [12] | 68.99 | 62.75 | 5 seconds |
| Aubio [13] | 57.09 | 55.91 | 3 consistent beat periods |
| | Offline beat tracking methods | | |
| CF [24] | 67.97 | 67.74 | --- |
| DBN [25] | 77.75 | 77.36 | --- |

Figure 3 demonstrates the performance of DLB model on the GTZAN dataset for different number of particles. We can see that for greedy filtering setups such as N=300, DLB performs almost as well as the default setting. For very greedy setups such as N=100, it is still in the same performance range with state-of-the-art methods. The reason for the performance robustness against a low number of particles is that by using the efficient space state, positions and tempi can be represented with a smaller number of states. Also, contrary to the double/half tempo investigator's ineffectiveness in the high number of particles, they somewhat assist the performance in the settings with low number of particles, by adding a wider viewpoint to monitor.

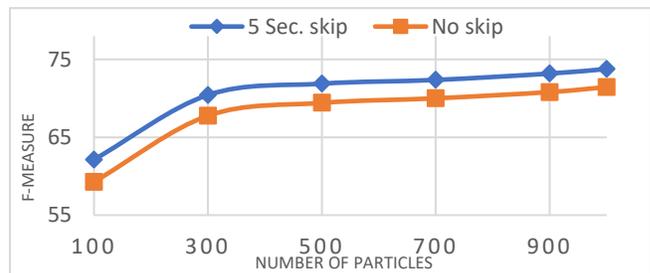

**Fig. 3:** DLB performance for different number of particles.

### 4. CONCLUSIONS AND FUTURE WORK

In this paper, we presented a novel online beat tracking method that leverages the beat activations of a unidirectional RNN in a sequential Monte Carlo particle filtering paradigm to infer music beat positions. By utilizing an optimized structure, we improved particle filtering performance in online beat tracking application. The proposed model is efficient and significantly outperforms state-of-the-art online beat tracking models. Also, we found that the performance of our casual inference model is close to that of non-casual state-of-the-art offline models. One direction for future work is to consider more rhythmic information in the model such as downbeat inference and meter analyses.